\documentclass[conference]{IEEEtran}
\IEEEoverridecommandlockouts
\usepackage{multirow}
\usepackage{booktabs}
\usepackage{cite}
\usepackage{amsmath,amssymb,amsfonts}
\usepackage{arydshln}
\usepackage{algorithmic}
\usepackage{graphicx}
\usepackage{textcomp}
\usepackage{xcolor}
\def\BibTeX{{\rm B\kern-.05em{\sc i\kern-.025em b}\kern-.08em
    T\kern-.1667em\lower.7ex\hbox{E}\kern-.125emX}}
\begin{document}

\title{SEE: Semantically Aligned EEG-to-Text Translation\\

\author{\IEEEauthorblockN{Yitian Tao\IEEEauthorrefmark{1}\thanks{Yitian Tao and Yan Liang contributed equally to this work.}, Yan Liang\IEEEauthorrefmark{1}, Luoyu Wang\IEEEauthorrefmark{1}, Yongqing Li\IEEEauthorrefmark{1}, Qing Yang\IEEEauthorrefmark{1}, and Han Zhang\IEEEauthorrefmark{1}\IEEEauthorrefmark{2}\IEEEauthorrefmark{3}\thanks{Corresponding author: Han Zhang }}
\IEEEauthorblockA{\IEEEauthorrefmark{1}School of Biomedical Engineering, ShanghaiTech University, Shanghai, China}
\IEEEauthorblockA{\IEEEauthorrefmark{2} State Key Laboratory of Advanced Medical Materials and Devices, ShanghaiTech University, Shanghai, China}
\IEEEauthorblockA{\IEEEauthorrefmark{3} Shanghai Clinical Research and Trial Center, Shanghai, China\\
zhanghan2@shanghaitech.edu.cn}

}}

\maketitle

\begin{abstract}
   % Electroencephalography (EEG), known for its non-invasiveness, ease of use, and cost-effectiveness, has been a popular method for capturing brain signals. However, current EEG-to-Text decoding approaches face challenges due to the huge domain gap between EEG recordings and raw texts, inherent data bias, and small closed vocabularies. In this paper, we propose SEE: Semantically Aligned EEG-to-Text Translation, a novel method aimed at improving EEG-to-Text decoding by seamlessly integrating two modules into a pre-trained BART language model: (1) a Cross-Modal Codebook that learns cross-modal representations to enhance feature consolidation and mitigate domain gap, and (2) a Semantic Matching Module that fully utilizes pre-trained text representations to align multi-modal features while considering noise caused by false negatives. Experimental results on the Zurich Cognitive Language Processing Corpus (ZuCo) demonstrate the effectiveness of SEE, significantly advancing the state of EEG-to-Text decoding.
   
   Decoding neurophysiological signals into language is of great research interest within brain-computer interface (BCI) applications. Electroencephalography (EEG), known for its non-invasiveness, ease of use, and cost-effectiveness, has been a popular method in this field. However, current EEG-to-Text decoding approaches face challenges due to the huge domain gap between EEG recordings and raw texts, inherent data bias, and small closed vocabularies. In this paper, we propose SEE: Semantically Aligned EEG-to-Text Translation, a novel method aimed at improving EEG-to-Text decoding by seamlessly integrating two modules into a pre-trained BART language model. These two modules include (1) a Cross-Modal Codebook that learns cross-modal representations to enhance feature consolidation and mitigate domain gap, and (2) a Semantic Matching Module that fully utilizes pre-trained text representations to align multi-modal features extracted from EEG-Text pairs while considering noise caused by false negatives, i.e., data from different EEG-Text pairs that have similar semantic meanings. Experimental results on the Zurich Cognitive Language Processing Corpus (ZuCo) demonstrate the effectiveness of SEE, which enhances the feasibility of accurate EEG-to-Text decoding.
\end{abstract}


\section{Introduction}
Decoding brain physiological signals to directly generate reading text is a rapidly emerging field in brain-computer interface (BCI) applications\cite{zhou2023speech2eeg, zhang2018converting, willett2021high, sun2020brain2char}, which is valuable for developing new communication methods for individuals with speech impairments or neuro-degenerative diseases \cite{sun2020brain2char}. It also provides deeper insights into the neural mechanisms of language processing, offering valuable perspectives on how the brain encodes and decodes linguistic information \cite{pfeiffer2020neuroimage}. In this field, electroencephalography (EEG) is widely used due to its cost-effectiveness, ease of use, non-invasiveness, and insensitivity to motion artifacts. These advantages facilitate paradigm design and data acquisition compared to electrocorticography (ECoG) and functional magnetic resonance imaging (fMRI) \cite{xu2021EEGBCIreview,varbu2022eegadvantagetofmri,liu2024eegadvantagetoEcoG}. The Zurich Cognitive Language Processing Corpus (ZuCo) was created using EEG \cite{hollenstein2018zuco1.0}. By mapping EEG signals during natural text reading to the semantic and syntactic elements of language, text generation based on real-time brain activity—known as EEG-to-Text decoding—can be achieved.

EEG-to-Text decoding has made significant progress, yet it remains constrained by limitations in vocabulary size and poor semantic understanding ability caused by a vast EEG-Text domain gap. Early studies \cite{saha2019speak,nieto2022thinking,moses2021neuroprosthesis} achieved high accuracy with small, closed vocabularies, focusing on recognizing low-level linguistic features such as individual words or syllables. However, these methods struggled to capture complex, high-level sentence and context information, making them unsuitable for open-vocabulary tasks. The development of large language models (LLMs) has advanced the field, with models like BART \cite{lewis2019bart} being adapted for EEG-based decoding \cite{wang2022preprocessing,liu2024eeg2text}. Wang and Ji were among the first who incorporated an EEG encoder with BART by aligning EEG recordings with pre-trained language models \cite{wang2022preprocessing}. Recent studies have expanded vocabulary size dramatically, from hundreds to tens of thousands of words, yet challenges remain in bridging the domain gap between EEG recordings and text since the predicted texts are sometimes irrelevant compared with ground truth text. To address this challenge, methods like contrastive learning\cite{chen2020simple, radford2021learning} were applied to enhance the quality of EEG-to-Text decoding through cross-modal alignment, which pulls the representation extracted from the same EEG-Text pairs together and pushes others apart \cite{feng2023aligning,duan2024dewave}. However, due to the impact of the noise caused by false negatives (i.e., data from different EEG-Text pairs that have similar semantic meanings) and the inherent data bias, the performance of these algorithms is still far from expectation. Recent research, such as MedCLIP \cite{wang2022medclip}, tried to mitigate the impact of false negatives using labels such as diagnosed diseases, while in EEG-to-Text translation task, no direct labels can be used for additional supervision.

In this paper, we propose SEE, a Semantically Aligned EEG-to-Text Translation method considering both the interaction between EEG recordings and their corresponding texts yet mitigating the influence of false negative EEG-Text pairs. SEE consists of two carefully designed modules seamlessly embedded into a pre-trained language model BART \cite{lewis2019bart}: 1) A Cross-Modal Codebook that learns cross-modal shared representations during the training period, thus suggesting feature consolidation and modality bias mitigation which helps to translate EEG to Text more easily; 2) A Semantic Matching module which is capable of aligning multi-modal features while considering the semantic consistency of false negative pairs by fully exploiting the text representations produced by pre-trained language model. We test our model on the ZuCo dataset, and experimental results show the superiority of our proposed method.

The contributions of this work include:
\begin{itemize}
    \item We design a Cross-Modal Codebook for learning cross-modal mutual representations during the training period. By consolidating features during inference, we can mitigate the domain gap and ease the difficulty of EEG-to-Text translation.
    \item A Semantic Matching module is adopted to ensure cross-modal semantic consistency while mitigating the impact of false negatives, leading to better cross-modal interaction.
    \item We seamlessly integrate the above two well-designed modules into a pre-trained language model, leveraging our prior knowledge of language modeling. This kind of integration is demonstrated to be important through the evaluation on the ZuCo dataset, which presents State-of-the-art results.
\end{itemize}

\section{Method}

\begin{figure*}[h]
  \centering
  \includegraphics[width=0.9\linewidth]{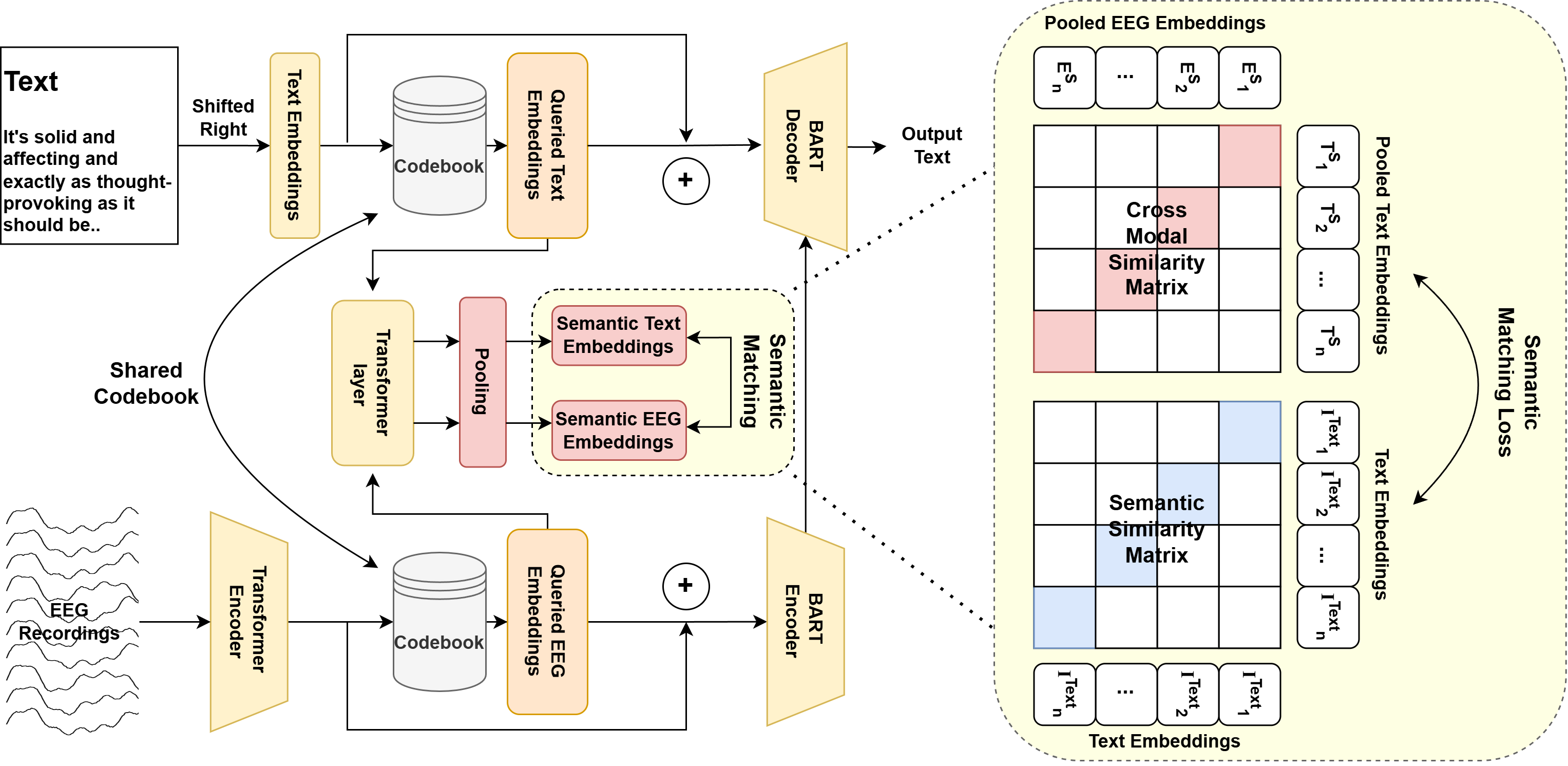}
  \caption{Illustration of SEE, where two modules are seamlessly embedded into a pre-trained transformer-based language model BART as a whole: 1) A Cross-Modal Codebook M which stores cross-modal representations for multi-modal retrieval, thus suggesting feature enhancement and modality bias mitigation; 2) A Semantic Matching module which is capable of aligning multi-modal features while considering the semantic consistency.}
  \label{figure:1}
\end{figure*}

The overall model structure of our proposed SEE model is depicted in Figure \ref{figure:1}. For better leveraging the inherent cross-modal semantic consistency while bridging the gap between different modalities for better EEG-to-Text translation, we design two modules seamlessly embedded into a pre-trained transformer-based \cite{vaswani2017attention} language model BART \cite{lewis2019bart} to fully harness the prior knowledge of language modeling: 1) A Cross-Modal Codebook $M$ which stores cross-modal representations for multi-modal retrieval, thus suggesting feature enhancement and modality bias mitigation; 2) A Semantic Matching module which is capable of aligning multi-modal features while considering the semantic consistency (i.e., the problem of false negative pairs). 
Similar to language modeling tasks such as image captioning \cite{tao2024memory, chen2022cross}, the EEG-to-Text translation task can be viewed as maximizing the probability of generating texts conditioned on the EEG recording $E$ and Codebook $M$:
\begin{equation}
    p(T|E)=\prod_{t=1}^l p(T_t|T_1,T_2,...,T_{t-1},f(E), M),
\end{equation}
where $T$ is the target text, $l$ is the text length, and $f(.)$ is our proposed model. 

\subsection{Cross-Modal Codebook Retrieval}
During training, a learnable shared Codebook $M\in\mathbb{R}^{N_C \times N_d}$ (where $N_C$ is the size of the Codebook and $N_d$ is the dimension of representations) is designed for learning shared cross-modal information from EEG-Text pairs which can be used for cross-modal feature consolidation, thus making it possible that additional information can be queried and integrated during inference (when the EEG recordings are the only available input). To achieve that, for each preprocessed ZoCo EEG recording $E$, we use an additional transformer encoder to extract EEG features $E^f\in\mathbb{R}^{l_E \times N_d}$ (where $l_E$ is the length of a preprocessed EEG recording). As for its corresponding text $T$, we adopted the pre-trained word embeddings of BART model: $\textbf{S}=\{\textbf{t}_1, \textbf{t}_2, \textbf{t}_3,...,\textbf{t}_l \}$. Then, the cross-modal Codebook retrieval process allows representations of different modalities to retrieve the Codebook through a cross-attention mechanism\cite{vaswani2017attention}:
\begin{equation}
    Q^E=EW^Q, \quad Q^S=SW^Q, \quad K=MW^K,
\end{equation}
\begin{equation}
    J^E = Softmax\left(\frac{Q^EK}{\sqrt{N_d}}\right),\quad J^S=Softmax\left(\frac{Q^SK}{\sqrt{N_d}}\right),
\end{equation}
where $W^Q$ and $W^K$ are projection matrices, $J^E$ and $J^S$ represent the similarity maps of EEG and Text modalities for querying Codebook, respectively.
Following the approach in \cite{chen2022cross, tao2024memory}, for each single-modality representation vector $E^f_i$ or $\textbf{t}_i$, we select the top $k$ elements from the Codebook that have the highest similarity scores with the representation vector. The queried multi-modal embeddings $P^E$ and $P^S$ are retrieved from the Codebook by computing a weighted average with a \textit{Softmax} function applied to the similarity scores:
\begin{equation}
    P^E=Softmax(C^E_{sim})(M^E_{retrieved}),
\end{equation}
\begin{equation}
    P^S=Softmax(C^S_{sim})(M^S_{retrieved}),
\end{equation}
where $M^E_{retrieved}$, $M^S_{retrieved}$, $C^E_{sim}$, and $C^S_{sim}$ represent the top $k$ elements along with their corresponding similarity scores for the EEG and text representations, respectively.
To integrate the queried multi-modal embeddings into the text generation process, we add it to the corresponding modality representations, obtaining the fine-grained consolidated features:
\begin{equation}
    E' = E^f + P^E, \quad S' = S + P^S.
\end{equation}

\subsection{Semantic Matching}
After cross-modal Codebook retrieval, we adopted a semantic matching method for aligning multi-modal representations to ensure the consistency of queried embeddings. A shared additional transformer layer $f_{trans}(.)$ along with average pooling layer $f_{Pool}(.)$ are first utilized to obtain semantic representations for EEG and Text modalities, respectively, from the queried multi-modal embeddings:
\begin{equation}
    E^S = f_{Pool}(f_{trans}(P^E)),\quad T^S = f_{Pool}(f_{trans}(P^S)).
\end{equation}
Then, considering a batch of multi-modal semantic representations $E^S\in\mathbb{R}^{N_{batch} \times N_d}$ and $T^S\in\mathbb{R}^{N_{batch} \times N_d}$, we can calculate the cross-modal similarity matrix $I=E^S (T^S)^T$. However, if we directly adopt contrastive loss\cite{chen2020simple} to push the semantic representations from the same EEG-Text pairs together (i.e., maximize the diagonal value of $I$) and others apart (i.e., minimize the off-diagonal value of $I$), huge noise will be introduced because of false negative pairs \cite{wang2022medclip}. Since the raw texts themselves produce enough semantic meanings, they can represent the meaning of their corresponding EEG-Text pairs and help recognize false negative pairs. To exploit this characteristic, we use a parameter-frozen pre-trained BART encoder, which has learned prior knowledge of language modeling, to encode all the texts in a batch and leverage average pooling to project each text into $N_d$ dimension vectors which maintain rich semantic information: $I^{Text}\in\mathbb{R}^{N_{batch} \times N_d}$. We can get soft labels indicating the semantic similarities of all EEG-Text pairs by calculating:
\begin{equation}
    \hat{I}_{ij} = \frac{I^{Text}_i(I^{Text}_j)^T}{||I^{Text}_i|| \times ||I^{Text}_j||}.
\end{equation}
After that, we evaluate the soft label matrix $\hat{I}$ and use a dynamic weighting function $D(x) = 1-x $ to mask the off-diagonal values that are over a specific threshold $\alpha$ (which is set to 0.5 by default):

\begin{equation}
    \hat{I}_{ij} = 
    \begin{cases} 
    \hat{I}_{ij}, & \text{if } \hat{I}_{ij} < \alpha, \\
    D(\hat{I}_{ij}), & \text{otherwise.}
    \end{cases}
\end{equation}

Finally, similar to MedCLIP \cite{wang2022medclip}, a \textit{Softmax} function is adopted on $\hat{I}$ to get:
\begin{equation}
    \hat{I}'_{ij} = \frac{exp \ \hat{I}_{ij} }{\sum_{j=1}^{N_{batch}}exp \ \hat{I}_{ij}},
\end{equation} 
and the semantic matching loss can be formulated as:
\begin{equation}
    L_{semantic\_matching} = - \frac{1}{N_{batch}}\sum_{i=1}^{N_{batch}}\sum_{j=1}^{N_{batch}}\hat{I}_{ij}’\log I_{ij}.
\end{equation}
By doing so, the contribution of those false negatives in the loss function is mitigated, and the noise can be minimized.

\subsection{Text Decoding}
For decoding a single word $s_u$ at time step $u$, the fine-grained consolidated features $E'$ and $S'$ (of previous time steps) are fed into the BART decoder $f_{decoder}$ (.):
\begin{equation}
    s_u=f_{decoder}(E', S_1',S_2',....S_{u-1}').
\end{equation}

Then, the report generation loss can be formulated as a cross-entropy loss:
\begin{equation}
    L_{gen}=-\frac{1}{l}\sum_{i=1}^l \sum_{j=1}^{V_d}T_{ij}log(s_{ij}),
\end{equation}
where $l$ is the length of the report, $V_d$ is the vocabulary size, $T_{ij}$ and $s_{ij}$ are the $j_{th}$ element of ground truth one-hot vector of the $i_{th}$ word and the predicted word, respectively. 

Collectively, our model is trained by minimizing the joint loss consisting of both report generation loss and semantic alignment contrastive loss :

\begin{equation}
    L = L_{gen} + L_{semantic\_matching}.
\end{equation}

Due to the use of fine-grained consolidated features that derived from the Codebook, the multi-modal information exchange of the Codebook is maximized, which contributes to better optimization.

\section{Experiments}
\begin{table*}[htbp]
  \caption{Examples of EEG-to-Text decoding using SEE.}
  \label{table:1}
  \centering
  \begin{tabular}{cc}
    \toprule

    \multirow{2}{*}{1} & Ground Truth: The sort of movie that gives tastelessness a bad rap.\\
     & Predicted: movie of movie that will youeless, a bad name.   \\
     \midrule
    \multirow{2}{*}{2} & Ground Truth: Bray is completely at sea; with nothing but a Savage Garden music video on his resume, he has no clue about making a movie. \\
     & Predicted: Rob a unf sea. his the to the few hook to and to the phone. and spends no immediate what contemporary a movie\\
     \midrule
    \multirow{2}{*}{3} & Ground Truth: It's not a particularly good film, but neither is it a monsterous one.\\
     & Predicted: 's not a good romantic movie, but it is it a masterpiece. one.\\
    \midrule
    \multirow{2}{*}{4} & Ground Truth: This odd, poetic road movie, spiked by jolts of pop music, pretty much takes place in Morton's ever-watchful gaze \\
     & Predicted: is, mood comedy which with theumboting of energy music, has much guarantees you in the's world-widful brain. \\
    \bottomrule

  \end{tabular}
\end{table*}
\subsection{Dataset and Metrices}
For our study, we use the ZuCo 1.0 \cite{hollenstein2018zuco1.0} and ZuCo 2.0 \cite{hollenstein2019zuco2.0} datasets, which provide EEG and eye-tracking data collected from healthy native English-speaking adults during five different English natural reading tasks. ZuCo 1.0 includes two normal reading tasks (SR v1.0 and NR v1.0) and one task-specific reading task (TSR v1.0). SR v1.0 uses movie reviews with sentimental content, while the other tasks use Wikipedia text. As for ZuCo 2.0, we use only NR v2.0, also based on Wikipedia. Word-level EEG data were aligned with eye-tracking data, following preprocessing steps and dataset splits from previous work \cite{wang2022preprocessing}.

To evaluate the performance of our proposed method, we choose natural language generation metrics following \cite{wang2022preprocessing}, which are BLEU-4 score \cite{papineni2002bleu} and ROUGE-1 scores \cite{lin2004rouge}.

As for the implementation details, we use a 3-layer Transformer encoder to extract EEG features and set the Codebook size to 1024. The learning rate is set to $5\times10^{-4}$, and the batch size is 32. The model is trained on an RTX4090 GPU using Adam as the optimizer.

\subsection{Experimental Analysis}

\begin{table}[htbp]
  \caption{Comparison with previous studies}
  \label{table:2}
  \centering
  \begin{tabular}{ccccccc}
    \toprule
     Method & BLEU-4 & ROUGE-P & ROUGE-R & ROUGE-F \\
    \midrule
     Transformer \cite{vaswani2017attention} & 0.09 & 12.6 & 13.5 & 12.5 \\
     BART \cite{lewis2019bart} & 6.3 & 31.5 & 29.8 & 30.5 \\
     EEG-to-Text \cite{wang2022preprocessing} & 6.8 & 31.6 & 28.8 & 30.1 \\
     SEE (ours) & \textbf{7.7} & \textbf{32.3}  & \textbf{29.9} & \textbf{31.1} \\
    \bottomrule

  \end{tabular}
\end{table}

% \begin{table*}[h]
%   \caption{Ablation studies of SEE. The BASE model is a pretrained BART model. Codebook and Semantic matching denote the use of the Codebook and the Semantic Matching, respectively.}
%   \label{table:1}
%   \centering
%   \begin{tabular}{ccccccccc}
%     \toprule
%     Trial & Method & BLEU-4 & ROUGE-1-P & ROUGE-1-R & ROUGE-1-F & ROUGE-L-P & ROUGE-L-R & ROUGE-L-F \\
%     \midrule
%     1 & BASE & 6.3 & 32.3 & 29.9 & 31.1 & 29.1 & 26.9 & 27.9 \\
%     2 & BASE + Codebook & 7.0 & 32.9 & 30.0 & 31.0 & 28.8 & 27.0 & 27.8 \\
%     3 & BASE + Semantic Matching & 7.2 & 31.6 & 29.4 & 30.4 & 28.6 & 26.7 & 27.6\\
%     4 & SEE & 7.7 & 31.5 & 29.8 & 30.5 & 28.7 & 27.2 & 27.9 \\
%     \bottomrule

%   \end{tabular}
% \end{table*}
\begin{table}[htbp]
  \caption{Ablation studies of SEE. The BASE model is a pretrained BART model. Codebook and Semantic denote the use of the Codebook and the Semantic Matching, respectively.}
  \label{table:3}
  \centering
  \begin{tabular}{cccccccc}
    \toprule
     Method & BLEU-4 & ROUGE-P & ROUGE-R & ROUGE-F \\
    \midrule
     BASE & 6.3 & 31.5 & 29.8 & 30.5\\
     BASE + Codebook & 7.0 & \textbf{32.9} & \textbf{30.0 }& 31.0 \\
     BASE + Semantic & 7.2 & 31.6 & 29.4 & 30.4 \\
     SEE (from scratch) & 4.9& 22.0 &21.5 & 21.6\\
     SEE & \textbf{7.7} & 32.3  & 29.9 &\textbf{ 31.1 } \\
    \bottomrule

  \end{tabular}
\end{table}

Table \ref{table:1} shows the examples of EEG-to-Text decoding using SEE model. From Table \ref{table:2}, we observe that the SEE model achieves superior performance compared to both the baseline BART model and the EEG-to-Text model\cite{wang2022preprocessing} across all metrics, especially in BLEU-4 score. These results indicate that by integrating both Codebook and Semantic Matching, our SEE model produces higher-quality and fluent text generation from EEG data than other competing models. 

In Table \ref{table:3}, the performance of the SEE model is further explored through ablation studies to analyze the individual contributions of the Codebook and Semantic Matching mechanisms. 
The SEE model, which combines both codebook and semantic matching, yields the best results. This demonstrates that the combination of both mechanisms leads to the most balanced performance in terms of both precision and recall, ultimately generating more coherent and accurate text from EEG recordings. We can also see that, without the pre-trained language model (a transformer model trained from scratch), the performance of the model drops drastically, which implies the importance of integrating the pretrained language model in EEG-to-Text translation.

\textbf{The effect of Codebook:} Adding the Codebook increases the BLEU-4 score and significantly boosts the precision of ROUGE-1, showing that the codebook helps mitigate domain bias between EEG recordings and text outputs. This improvement underscores the importance of capturing fine-grained features between EEG signal patterns and text elements.

\textbf{The effect of Semantic Matching:} Introducing semantic matching improves both the BLEU-4 score (7.2) and the ROUGE-R score (29.4), suggesting that semantic alignment between EEG features and text generation helps ensure the cross-modal consistency, which lays the foundation of exploring the deep correlation between EEG and Text modalities and contributes to smooth EEG-to-Text translation. However, the ROUGE-F score only marginally increases to 30.4, indicating that semantic matching alone might not be sufficient for optimal performance.

\section{Conclusion}
In this paper, we presented Semantically Aligned EEG-to-Text Translation (SEE), a novel approach that addresses the challenges in EEG-to-Text decoding, particularly the cross-modal domain gap and data bias that hinder current methods. By integrating a cross-modal Codebook and a Semantic Matching Module into a pre-trained BART language model, SEE enhances cross-modal representation learning and aligns multi-modal features with greater precision, even accounting for noise due to false negatives. Our experiments on the ZuCo dataset validate the effectiveness of SEE, showing significant improvements in EEG-to-Text decoding accuracy compared to other methods which directly translate EEG recordings to text and ignore cross-modal dependency.

% This study presents an innovative research framework including simultaneous functional PET/MR, multimodal brain connectome construction and learning, clinically feasible multimodal fusion and diagnosis. The superior performance of MX-ARM is demonstrated by a precious, carefully curated sf-PET/MR dataset. The AUC of 0.827 also outperforms current SOTA performance in MCI detection, indicating that concurrent modeling brain metabolic, hemodynamic, and perfusion activity helps with more accurate early AD detection.

\section*{Acknowledgment}
This work is partially supported by the STI2030-Major Project (No. 2022ZD0209000) and the Shanghai Pilot Program for Basic Research - Chinese Academy of Science, Shanghai Branch (No. JCYJ-SHFY-2022-014).

\bibliographystyle{ieeetr}
\bibliography{ref}

\end{document}